\documentclass[12pt]{article}
\pdfoutput=1
\usepackage{draft,tikz-cd}

\newcommand{\PSU}{\text{PSU}}
\newcommand{\SU}{\text{SU}}
\newcommand{\U}{\text{U}}

\begin{document}

\begin{titlepage}

\begin{center}

\hfill \\
\hfill \\
\vskip 1cm

\title{Words to describe a black hole}

\author{Chi-Ming Chang$^{a,b}$ and Ying-Hsuan Lin$^{c}$}

\address{\small${}^a$Yau Mathematical Sciences Center (YMSC), Tsinghua University, Beijing 100084, China}

\address{\small${}^b$Beijing Institute of Mathematical Sciences and Applications (BIMSA), Beijing 101408, China}

\address{${}^c$Jefferson Physical Laboratory, Harvard University, Cambridge, MA 02138, USA}

\email{cmchang@tsinghua.edu.cn, yhlin@fas.harvard.edu}

\end{center}

\vfill

\begin{abstract}
    We revamp the constructive enumeration of 1/16-BPS states in the maximally supersymmetric Yang-Mills in four dimensions, and search for ones that are not of multi-graviton form.  A handful of such states are found for gauge group SU(2) at relatively high energies, resolving a decade-old enigma.  Along the way, we clarify various subtleties in the literature, and prove a non-renormalization theorem about the exactness of the cohomological enumeration in perturbation theory.  We point out a giant-graviton-like feature in our results, and envision that a deep analysis of our data will elucidate the fundamental properties of black hole microstates.
\end{abstract}

\vfill

\end{titlepage}

\tableofcontents

\section{Introduction}

Recently, the Event Horizon Telescope \cite{EventHorizonTelescope:2019ths} astounded the world with the first image of a black hole at the center of our galaxy.  Instead of pictures, we search for ``words'' to describe black hole microstates.
The seminal work of Strominger and Vafa \cite{Strominger:1996sh} matching the area of a black hole horizon with the counting of microstates has fundamentally changed our perception of string theory and quantum gravity.
Not only did it fortify the legitimacy of string theory as a consistent theory of quantum gravity, but it also revealed that black hole horizons contain a large amount of quantum information, contrary to the classical no-hair theorem.
Extracting and decoding this information is key to unveiling profound fundamental aspects of black holes.
While \cite{Strominger:1996sh} sparked bustling activity in \emph{counting} black hole microstates, much less effort has been invested in uncovering the properties of the states themselves.

The AdS/CFT duality provides a rigorous framework to attack this problem. The Bekenstein-Hawking entropy of an asymptotically-AdS black hole is dual to the statistical entropy of a thermal state in the conformal field theory (CFT) living at the boundary of the AdS space.  In the classic correspondence between the type IIB superstring theory on ${\rm AdS}_5\times {\rm S}^5$ and the ${\cal N}=4$ super-Yang-Mills (SYM) theory with gauge group $\SU(N)$, the entropy of electrically-charged rotating BPS black holes in AdS$_5$ \cite{Gutowski:2004ez,Gutowski:2004yv,Chong:2005hr,Chong:2005da,Kunduri:2006ek} was expected to be reproduced by the superconformal index of the CFT \cite{Kinney:2005ej}. However, due to large cancellations between bosonic and fermionic states at nearby charges, the superconformal index at real saddle points is not large enough
to account for the growth of the number of black hole microstates. It was then realized in \cite{Cabo-Bizet:2018ehj}, building on prior observation by \cite{Hosseini:2017mds}, that the black hole entropy is related to a complexified Euclidean on-shell action. The complexified bulk geometry identified a set of complex fugacities, and the superconformal index of the CFT with such fugacities exactly reproduced the black hole entropy \cite{Cabo-Bizet:2018ehj,Choi:2018hmj,Benini:2018ywd}.\footnote{Earlier, the topologically twisted indices in CFT$_3$ reproduce the entropy of a class of static dyonic BPS black holes in AdS$_4$ \cite{Benini:2015eyy,Benini:2016rke}. Generalizations to other dimensions and other amounts of supersymmetry can be found in
\cite{Choi:2018vbz,Honda:2019cio,Fluder:2019szh,ArabiArdehali:2019tdm,Kim:2019yrz,
Gauntlett:2019roi,Hosseini:2019ddy,Kim:2019umc,Cabo-Bizet:2019osg,
Amariti:2019mgp,Gang:2019uay,Larsen:2019oll,Kantor:2019lfo,GonzalezLezcano:2019nca,Lanir:2019abx,Choi:2019zpz,Bobev:2019zmz,Nian:2019pxj,Cabo-Bizet:2019eaf,Benini:2019dyp,Goldstein:2019gpz,ArabiArdehali:2019orz
,David:2020ems,Cabo-Bizet:2020nkr,Murthy:2020rbd,Agarwal:2020zwm,Benini:2020gjh,GonzalezLezcano:2020yeb,Copetti:2020dil,Goldstein:2020yvj,Cabo-Bizet:2020ewf,Amariti:2020jyx,Choi:2021lbk,Amariti:2021ubd,Jejjala:2021hlt,Cassani:2021dwa
,Ezroura:2021vrt}.}

In light of these triumphs, it is natural to take one step further and try to understand not just the number of black hole microstates, but to find their precise holographic dual, which by the state/operator correspondence are local operators in ${\cal N}=4$ SYM. 
This problem has two parts:
\begin{enumerate}
    \item  {\it Enumeration of BPS operators}. This has been studied at weak coupling in \cite{Berkooz:2006wc,Janik:2007pm,Grant:2008sk,Chang:2013fba}, by
    organizing them into the cohomology of a supercharge $Q$ on the Fock space of single-trace and multi-trace operators.\footnote{\label{ft:hodge_argument}While the $Q$-closed condition is natural, the unfamiliar reader may at this point wonder about the meaning of $Q$-exactness.  A 1/16-BPS operator must also be annihilated by $Q^\dag$, which receives quantum corrections.  To circumvent this difficulty, one invokes Hodge theory which establishes a bijection between harmonic forms (annihilated by $\Delta\equiv2\{Q, Q^\dag\}$) and the de Rham cohomology ($Q$-cohomology). Note that $Q$-exact operators are orthogonal to the 1/16-BPS operators, and $Q$-closed operators with non-zero $\Delta$-eigenvalues are $Q$-exact.}
    The weak-coupling spectrum was then conjectured by \cite{Grant:2008sk} to be invariant along the exactly marginal deformation into the strong-coupling regime, which is appropriate for describing black holes.  In \cite{Chang:2013fba}, the $Q$-cohomology was reformulated as a relative Lie superalgebra cohomology.
    \item  {\it Identification of the ones dual to black holes,} and not just a gas of gravitons.  To this end, explicit representatives of cohomology classes dual to the multi-gravitons were proposed in \cite{Chang:2013fba} and shown to recover the multi-graviton partition function in the AdS at large $N$.  In \cite{Chang:2013fba}, a sporadic search at low ranks $N=2,3$ and reasonably high energies reported a negative result for the existence of states not of multi-graviton form, i.e.\ candidate black hole microstates.
    Nevertheless, the recent evaluation of finite $N$ indices strongly suggests that they should be present even at low ranks \cite{Murthy:2020rbd,Agarwal:2020zwm}.
\end{enumerate}

The present paper revisits this cohomology problem and performs a systematic search for cohomology classes that are not of multi-graviton form. Our main results can be summarized as follows:
\begin{enumerate}
    \item  We obtain a large collection of fully-refined counting data, which contains much more information than even the fully-refined index, let alone the unrefined index which most explicit evaluations consider.
    \item  The first non-graviton cohomology class is found for $N=2$ at energy $E=19/2$, disproving the conjecture of \cite{Chang:2013fba} about their nonexistence.  
    \item  A non-renormalization theorem is proven (assuming the applicability of Leibniz rule of the $Q$-action) in perturbation theory, which puts the conjecture of \cite{Grant:2008sk} on a firmer basis.
\end{enumerate}

The remainder of this paper is organized as follows.  Section~\ref{sec:letterandwords} introduces the BPS letters and words, reviews the formulation of the cohomology problem in terms of a BPS superfield, and discusses the BPS partition function that counts the BPS states (without signs). 
Section~\ref{sec:non-renormalization_theorem} proves the non-renormalization conjecture of \cite{Grant:2008sk} in perturbation theory, and comments on the non-perturbative extension.
The key results of our enumeration are then presented in Section~\ref{sec:results}, with various ramifications discussed.

\section{Letters and words}
\label{sec:letterandwords}

\subsection{Review of the cohomology problem}
\label{sec:review_coho}

The ${\cal N}=4$ super-Yang-Mills theory has superconformal symmetry $\PSU(2,2|4)$. Let us denote the 16 supercharges by $Q^I_\alpha$, $\overline Q_{I\dot\alpha}$, and the 16 superconformal supercharges by $S^\alpha_I$, $\overline S^{I\dot\alpha}$, where the upper (lower) $I=1,\cdots, 4$ is the (anti-)fundamental index of the $\SU(4)_R$ R-symmetry, and $\alpha,\dot\alpha=\pm,\dot\pm$ are the spinor indices of the Lorentz group $\text{SO}(4)\cong \SU(2)_L\times \SU(2)_R$. The 1/16-BPS operators are defined to be those annihilated by the supercharge $Q\equiv Q^4_-$ and its hermitian conjugate (BPZ conjugate) $Q^\dagger = S \equiv S^-_4$ in radial quantization. They have the commutator
\ie\label{eqn:one-loop_Hamiltonian}
\Delta\equiv 2\{Q,Q^\dagger\}=D-2J^3_L-q_1-q_2-q_3\,,
\fe
where $D$ is the dilatation operator, $J_L^3$ is the left $\SU(2)_L$ angular momentum, and $q_1$, $q_2$, $q_3$ are the Cartan generators of the ${\rm SO}(6)_R$ R-symmetry.\footnote{Our convention is that $R_1=q_2+q_3=R^4_4-R^1_1$, $R_2=q_1-q_2=R^1_1-R^2_2$, $R_3=q_2-q_3=R^2_2-R^3_3$, where $R_1$, $R_2$, $R_3$ are the Cartan generators of $\SU(4)_R$, and $R^1_1$, $R^2_2$, $R^3_3$, $R^4_4$ are the diagonal components of the fundamental representation of $\SU(4)_R$.}

By standard arguments (see footnote \ref{ft:hodge_argument}), the space of the 1/16-BPS operators is isomorphic to the cohomology of the supercharge $Q$ \cite{Grant:2008sk}. In the weak coupling limit, one can further restrict the $Q$-cohomology to the classical null space of  $\Delta$.
In the path integral formalism, local operators are constructed out of gauge-invariant combinations of the fundamental fields, which consist of six scalars $\Phi_{IJ}$ in the antisymmetric representation of $\SU(4)_R$ with the reality condition $\Phi_{IJ}^*=\frac{1}{2}\epsilon^{IJKL}\Phi_{KL} \equiv\Phi^{IJ}$, four chiral fermions $\Psi_{I\alpha}$ as well as their complex conjugates $\overline \Psi^{I\dot \alpha}$, and the gauge field $A_\mu$.  

The operators in the classical null space of  $\Delta$ are constructed using the BPS \emph{letters}, which are fundamental fields and their derivatives of vanishing classical $\Delta$-eigenvalue.\footnote{This is also referred as the ${\rm psu}(1,2|3)$ subsector \cite{Harmark:2007px}.} 
The full set of BPS letters are
\ie\label{eqn:BPS_letters_1}
\phi^i\equiv \Phi^{4i}\,,\quad \psi_i\equiv -i\Psi_{i+}\,,\quad\lambda_{\dot\alpha}\equiv \bar\Psi^4_{\dot\alpha}\,,\quad f\equiv-iF_{++}= F_{\mu\nu}(\sigma^{\mu\nu})_{++}\,,
\fe
where $i=1,\,2,\,3$, and their covariant $D_{\dot\alpha}$-derivatives, where
\ie\label{eqn:BPS_letters_2}
D_{\dot\alpha}\equiv  (\sigma^\mu)_{+\dot\A}D_\mu\,.
\fe
The BPS letters satisfy two relations. The first is the obvious relation between $D_{\dot\alpha}$ and $f$,
\ie\label{eqn:relation_1}
[D_{\dot\alpha},D_{\dot\beta}]=\epsilon_{\dot\alpha\dot\beta} f\,.
\fe
The second is the only equation of motion that is purely made out of the BPS letters,
\ie\label{eqn:relation_2}
D_{\dot\alpha}\lambda^{\dot\alpha}=[\phi^i,\psi_i]\,,
\fe
which amounts to the invariance of the path integral under field redefinitions.
The gauge-invariant combinations of the BPS letters \eqref{eqn:BPS_letters_1} and \eqref{eqn:BPS_letters_2} up to the relations \eqref{eqn:relation_1} and \eqref{eqn:relation_2} will be referred as the BPS \emph{words}. The supercharge $Q$ acts on the BPS letters as
\ie
\relax [Q,\phi^i]&=0\,,\quad \{Q,\psi_i\}=-i\epsilon_{ijk}[\phi^j,\phi^k]\,,
\\
\{Q,\lambda_{\dot\alpha}\}&=0\,,\quad [Q,f]=i[\phi^i,\psi_i]\,,
\\
[Q,D_{\dot \alpha}\zeta]&=-i[\lambda_{\dot\alpha},\zeta]+D_{\dot \alpha}[Q,\zeta]\,,&
\fe
where $\zeta$ is any fundamental field.

The BPS fields can be assembled into a fermionic ``BPS superfield" \cite{Chang:2013fba}, which is a power series in the Grassmann variables $\theta_i$,
\ie\label{eqn:BPS_superfield}
\Psi(z,\theta)=-i[\lambda(z)+2\theta_i\phi^i(z)+\epsilon^{ijk}\theta_i\theta_j\psi_k(z)+4\theta_1\theta_2\theta_3f(z)]\,,
\fe
and the component fields $\lambda(z)$, $\phi^i(z)$, $\psi_i(z)$, $f(z)$ are formal power series of the auxiliary variables $z^{\dot\alpha}$ \cite{Grant:2008sk},
\ie\label{eqn:BPS_fields}
\phi^i(z)&=\sum^\infty_{n=0}\frac{1}{n!}(z^{\dot \alpha} D_{\dot\alpha})^n \phi^i\,,& \psi_i(z)&=\sum_{n=0}^\infty\frac{1}{n!}(z^{\dot\alpha}D_{\dot\alpha})^n\psi_i\,,
\\
\lambda(z)&=\sum^\infty_{n=0}\frac{1}{(n+1)!}(z^{\dot \alpha} D_{\dot\alpha})^n (z^{\dot\beta}\lambda_{\dot\beta})\,,&f(z)&=\sum^\infty_{n=0}\frac{1}{n!}(z^{\dot \alpha} D_{\dot\alpha})^n f\,.
\fe
The BPS superfield satisfies the constraint
\ie
\Psi({\cal Z}=0)=0\,,
\fe
where we collectively denote all the auxiliary variables as ${\cal Z}=(z^+,z^-;\theta_1,\theta_2,\theta_3)$. The BPS letters can be recovered by taking the $\partial_{z^\pm}$ and $\partial_{\theta_i}$ derivatives of the BPS superfield $\Psi$ and evaluating it at the origin ${\cal Z}=0$ of the superspace $\bC^{2|3}$. One does not miss anything by symmetrizing all the $\dot\alpha$ indices in \eqref{eqn:BPS_fields} because the antisymmetric parts can be replaced by the right-hand sides of \eqref{eqn:relation_1} and \eqref{eqn:relation_2}. 
The charges of the BPS superfield and the derivatives are given in Table \ref{tab:charges}, where $Y$ is related to the charge of the ``bonus" ${\rm U}(1)_Y$ symmetry \cite{Intriligator:1998ig,Intriligator:1999ff}.\footnote{Compared with \cite[(5.2)]{Intriligator:1998ig}, $Y_\text{there} = - 2 Y_\text{here} + 2 \sum_{i=1}^3 q_i$.
}
\begin{table}[H]
\begin{center}
\begin{tabular}{|c|c|c|c|c|}
\hline
Charges & $J^3_L$ & $J^3_R$ & $q_j$ & $Y$
\\
\hline\hline
$\Psi$ & $-\frac{1}{2}$ & 0 & $\frac{1}{2}$ & 1
\\
\hline
$\partial_{z^\pm}$ & $\frac{1}{2}$ & $\pm\frac{1}{2}$ & 0 & 0
\\
\hline
$\partial_{\theta_i}$ & $\frac{1}{2}$ & 0 & $\delta_{ij}-\frac{1}{2}$ & 0
\\\hline
\end{tabular}
\end{center}
\caption{\label{tab:charges} Angular momenta and charges of the BPS superfield and derivatives.
}
\end{table}

The supercharge $Q$ acts on the BPS superfield $\Psi({\cal Z})$ as
\ie\label{eqn:Q-action_on_Psi}
\{Q,\Psi({\cal Z})\}=\Psi({\cal Z})^2\,.
\fe
It was recognized in \cite{Chang:2013fba} that in this formulation the $Q$-cohomology is nothing but the relative Lie superalgebra cohomology 
\ie\label{eqn:relative_coh}
{\rm H}^*({\cal G}_N,{\rm sl}_N;{\mathbb C})\,,
\fe
where ${\cal G}_N\equiv {\mathbb C}[z^+,z^-]\otimes \Lambda[\theta_1,\theta_2,\theta_3]\otimes{\rm sl}_N$. For readers not familiar with Lie algebra cohomology, its basic definition can be found in Section~3 of \cite{Chang:2013fba}.

\subsection{BPS partition function and superconformal index}
\label{sec:BPS_partition_function}

The BPS partition function is defined as \cite{Chang:2013fba}
\ie
&Z(x,a,b,u,v,w)
\\
&={\rm Tr}_{{\cal H}_{\rm BPS}}\left[
x^Y a^{D-J_L^3+J_R^3-Y} b^{D-J_L^3-J_R^3-Y} u^{-q_2-q_3+Y} v^{-q_1-q_3+Y} w^{-q_1-q_2+Y}
\right],
\fe
where $x$ counts the number of the superfield $\Psi$, $a$ and $b$ count the number of the $z$-derivatives $\partial_{z_\pm}$, and $u$, $v$, $w$ count the number of $\theta$-derivatives $\partial_{\theta_i}$. 
The superconformal index defined in \cite{Kinney:2005ej} is given by specializing the BPS partition function as
\ie\label{eqn:SCI}
{I}_{\rm SCI}(t,y,v,w)&= Z(-1,t^3y,t^3/y,-t^2 v,-t^2w/v,-t^2 /w)
\\
&=\Tr\left[(-1)^F t^{2(D+J_L^3)}y^{2J_R^3}v^{R_2}w^{R_3}\right]\,,
\fe
where $R_1=q_2+q_3$, $R_2=q_1-q_2$, $R_3=q_2-q_3$, and the fermion number $F$ is the sum of the exponents of $x$, $u$, $v$, $w$, i.e.\ $
F=3Y-2q_1-2q_2-2q_3$. The unflavored index and BPS partition function are defined by
\ie
{I}_N(t)&\equiv{I}_{\rm SCI}(t,1,1,1)=Z(-1,t^3,t^3,-t^2,-t^2,-t^2)=\Tr\left[(-1)^F t^{2(D+J_L^3)}\right]\,,
\\
Z_N(t)&\equiv Z(1,t^3,t^3,t^2,t^2,t^2)=\Tr_{{\cal H}_{\rm BPS}}\left[ t^{2(D+J_L^3)}\right]\,.
\fe
Let us expand them as
\ie
{I}_N(t)=\sum_{n} d_N(n)t^n\,,\quad Z_N(t)=\sum_{n} {\bf d}_N(n)t^n\,,
\fe
where $n$ is the eigenvalue of $2(D+J_L^3)$.
The coefficients $d_N(n)$ and ${\bf d}_N(n)$ in the expansions count (up to signs for the case of $d_N(n)$) the BPS states.

There are two important asymptotic limits of the degeneracies $d_N(n)$ and ${\bf d}_N(n)$. First, in the $N\to\infty$ limit with $n$ fixed, the degeneracies only receive contributions from finite energy states, corresponding to (multi-)supergravitons in the AdS bulk. At large $n$, they behave as \cite{Kinney:2005ej}
\ie
\log |d_\infty(n)|&=\frac{\sqrt{5}\pi}{3}\sqrt{n}+O(\log n)\,,
\\
\log {\bf d}_\infty(n)&=\frac{2^\frac{1}{6}\sqrt{3}\pi}{5}n^\frac{5}{6}+O(\sqrt{n})\,.
\fe
Second, in the $N,n\to\infty$ limit with $j=n/N^2$ fixed, the degeneracies receive contributions from states with energy scaling as $N^2$, corresponding to black holes in the bulk. In this limit, the asymptotic behavior for $d_N(n)$ is \cite{Cabo-Bizet:2018ehj,Choi:2018hmj,Benini:2018ywd}
\ie
\lim_{N\to\infty} N^{-2}\log |d_N(N^2 j)|&=\frac{\pi}{2\cdot 3^\frac{1}{6}}j^\frac{2}{3}+O(j^\frac{1}{3})\,,
\fe
whereas the asymptotic behavior of ${\bf d}_N(n)$ remains an open problem.

For the index, $d_N(n)$ for ranks $N=2,\cdots,10$ were computed in \cite{Murthy:2020rbd,Agarwal:2020zwm}.
For the BPS partition function, we present in Section~\ref{sec:results} the results of ${\bf d}_N(n)$ for ranks $N=2,3,4$.\footnote{The ${\rm U}(N)$ and ${\rm SU}(N)$ BPS partition functions are simply related by
\ie
\frac{Z_{{\rm U}(N)}(x,a,b,u,v,w)}{Z_{{\rm SU}(N)}(x,a,b,u,v,w)}&=\exp\left[\sum_{n=1}^\infty \frac{1}{n}\left(Z_+(x^n,a^n,u^n,v^n,w^n)+(-1)^{n+1} Z_-(x^n,a^n,u^n,v^n,w^n)\right)\right]\,,
\\
Z_\pm(x,a,b,u,v,w)&=\frac{1}{2}\left(z(x,a,b,u,v,w)\pm z(- x,a,b,- u,- v,- w)\right)\,,
\\
z(x,a,b,u,v,w)&=\frac{(1+u)(1+v)(1+w)}{(1-a)(1-b)}x-x\,.
\fe
}
Note that $n$, defined as the eigenvalue of $2(D+J_L^3)$, is not to be confused with the \emph{energy} $E$, which is the eigenvalue of $D$.

\subsection{Multi-gravitons from single-trace cohomology}

The action \eqref{eqn:Q-action_on_Psi} of the supercharge $Q$ is closed in the space of single-trace BPS words. Hence, we can restrict the cohomology problem to this subspace, where all the single-trace $Q$-cohomology classes were found in \cite{Chang:2013fba}. They are represented by the single-graviton operators
\ie\label{eqn:graviton_op}
\partial^{p_1}_{z^+}\partial^{p_2}_{z^-}\partial^{q_1}_{\theta_1}\partial^{q_2}_{\theta_2}\partial^{q_3}_{\theta_3}\Tr\left[(\partial_{z^+}\Psi)^{k_1}(\partial_{z^-}\Psi)^{k_2}(\partial_{\theta_1}\Psi)^{m_1}(\partial_{\theta_2}\Psi)^{m_2}(\partial_{\theta_3}\Psi)^{m_3}\right]\Big|_{{\cal Z}=0}\,.
\fe
It is straightforward to check that \eqref{eqn:graviton_op} are $Q$-closed and also not $Q$-exact in the space of single-trace words.\footnote{Switching the ordering of the letters inside a trace results in operators in the same cohomology class.} The single-trace BPS partition function in the large $N$ limit was computed in \cite{Chang:2013fba}, which matched perfectly with the single-particle partition function in the bulk theory \cite{Kinney:2005ej}. 

Products of the single-graviton operators modulo trace relations represent non-trivial cohomology classes and are dual to multi-graviton states. We will refer to such cohomology classes as being \emph{of multi-graviton form}. The partition function counting the multi-gravition cohomology classes in the large $N$ limit is simply given by the plethystic exponential of the single-trace BPS partition function.

\section{A non-renormalization theorem}
\label{sec:non-renormalization_theorem}

The above counting of 1/16-BPS states is valid in the weak coupling limit. To compare with the entropy of gravitons and black holes in the bulk type IIB superstring theory, one needs to extend the counting to the regime of strong coupling. Let us analyze more precisely the validity regime of the counting in the coupling space using the relative Lie superalgebra cohomology \eqref{eqn:relative_coh}.
The generators of the Cartan subalgebra of the superconformal algebra all have integer or half-integer eigenvalues, except the dilatation operator $D$, and hence cannot receive quantum corrections. In the perturbation theory, the dilatation operator is expected to take the form
\ie
D=D_0+ \frac{g^2_{\rm YM}N}{8\pi^2}H+{\cal O}(g^3_{\rm YM})\,,
\fe
where $g_{\rm YM}$ is the gauge coupling, and $H$ is the one-loop Beisert Hamiltonian \cite{Beisert:2003jj,Beisert:2004ry}, which is a differential operator acting on the space of all gauge-invariant words. When restricting to the space of BPS words, $H$ is proportional to the anti-commutator of $Q$ and $Q^\dagger$, i.e.\footnote{Furthermore, $Q$ and $Q^\dagger$ at leading order in $g_{\rm YM}$ can be represented by
\ie\label{Qdagger}
Q=\Tr\left(\Psi \Psi\frac{\partial}{\partial \Psi}\right)\,,\quad
Q^\dagger=\Tr\left(\Psi\frac{\partial}{\partial \Psi}\frac{\partial}{\partial \Psi}\right)\,.
\fe}
\ie\label{eqn:H=QQ}
H\propto\{Q,Q^\dagger\}\,.
\fe
Hence, our discussion in the previous section shows that in the space of BPS words, the null space of the one-loop Beisert Hamiltonian $H$, is isomorphic to the cohomology ${\rm H}^*({\cal G}_N,{\rm sl}_N;{\mathbb C})$. 

It was conjectured in \cite{Grant:2008sk} that the supersymmetric spectrum of the ${\cal N}=4$ super-Yang-Mills theory on ${\rm S}^3$ is exactly given by the ground states of the one-loop Beisert Hamiltonian $H$ without any higher-loop or non-perturbative correction. Nontrivial checks of this conjecture by matching the one-loop partition function at infinite $N$ with the supergraviton partition function in the ${\rm AdS}_5\times {\rm S}^5$ was reported in \cite{Janik:2007pm,Chang:2013fba}. Let us prove the part of the conjecture of vanishing perturbative corrections.

The BPS state counting to all orders in perturbation theory is still governed by a relative Lie superalgebra cohomology.
We start by figuring out the relative cochain complex and the differential acting on it.
Firstly, BPS operators can be written as linear combinations of BPS words with coefficients in the formal power series ring $\bC[[g_{\rm YM}]]$. We do not need to include non-BPS words because operator-mixing in perturbation theory is only among those with the same angular momenta, R-charges, and classical dimension.\footnote{The dilatation operator $D$ obviously commutes with angular momenta and R-charges. In perturbation theory, $D$ also commutes with the classical dimension by dimensional analysis, and the classical dimensions of non-BPS words do not satisfy the BPS condition \eqref{eqn:one-loop_Hamiltonian}.}
Therefore, in terms of the BPS superfield $\Psi$, the BPS operators reside in the relative cochain complex ${\rm C}^*({\cal G}_N,{\rm sl}_N;\bC[[g_{\rm YM}]])$.

Next, it is important to note the advantage of our formulation that we only need the knowledge of the supercharge $Q$, but not the dilatation operator $D$. Unlike the dilatation operator which is deformed by the coupling $g_{\rm YM}$, one can choose a regularization scheme such that the action of $Q$ on the BPS superfield is undeformed perturbatively.\footnote{For example, one could use the ``dimensional regularization by dimensional reduction" (DRED) scheme \cite{Siegel:1979wq}, or the regularization scheme in \cite{Hayashi:1998ca}.}
Using the Leibniz rule, the $Q$-action on the BPS superfield \eqref{eqn:Q-action_on_Psi} gives the $Q$-action on the BPS words, or equivalently the differential acting on the relative cochain complex ${\rm C}^*({\cal G}_N,{\rm sl}_N;\bC[[g_{\rm YM}]])$. 
Here, we have assumed that the $Q$-action satisfies the Leibniz rule in perturbation theory.  We do not have a proof, but the only known obstruction, the Konishi anomaly \cite{Konishi:1983hf}, is absent in ${\cal N}=4$ SYM \cite{Kostyuk:2005iq}.\footnote{We thank Davide Gaiotto and Justin Kulp for pointing this subtlety out to us.}

In summary, the 1/16-BPS operators are classified by the relative Lie superalgebra cohomology
\ie
{\rm H}^*({\cal G}_N,{\rm sl}_N;\bC[[g_{\rm YM}]])\,,
\fe
which is isomorphic to the tensor product\footnote{The universal coefficient theorem implies
\[
    {\rm H}^*({\cal G}_N,{\rm sl}_N;\bC[[g_{\rm YM}]])\cong {\rm Hom}_\bC({\rm H}_*({\cal G}_N,{\rm sl}_N;\bC),\bC[[g_{\rm YM}]]).
\]
Viewing $\bC[[g_{\rm YM}]]$ as an infinite product $\bC\times \bC\times\cdots$ allows us to write
\[
    {\rm Hom}_\bC({\rm H}_*({\cal G}_N,{\rm sl}_N;\bC),\bC[[g_{\rm YM}]])\cong {\rm Hom}_\bC({\rm H}_*({\cal G}_N,{\rm sl}_N;\bC),\bC)\otimes_\bC \bC[[g_{\rm YM}]].
\]
Finally, we use the fact that
\[
    {\rm Hom}_\bC({\rm H}_*({\cal G}_N,{\rm sl}_N;\bC),\bC)\cong {\rm H}^*({\cal G}_N,{\rm sl}_N;\bC).
\]
We thank Dingxin Zhang for the discussion on this point.
}
\ie\label{eqn:coho_fac}
{\rm H}^*({\cal G}_N,{\rm sl}_N;\bC[[g_{\rm YM}]])\cong {\rm H}^*({\cal G}_N,{\rm sl}_N;\bC)\otimes_{\bC} \bC[[g_{\rm YM}]]\,.
\fe
Consequently, the 1/16-BPS spectrum is invariant to all orders in perturbation theory. Furthermore, by S-duality, our argument can be applied to all the weakly coupled points on the conformal manifold.

Finally, let us comment on the situation at finite couplings. Barring some subtleties, we expect that the BPS operators are linear combinations of BPS words with coefficients being holomorphic functions on the complex plane of the coupling $g_{\rm YM}$ at a finite distance away from the free points. While we do not know whether the cohomology still has an analogous factorization as \eqref{eqn:coho_fac}, the fact that the factorization exists at every weakly coupled point hints toward the affirmative.

\section{Constructive enumeration and discussions}
\label{sec:results}

\subsection{Computational scheme}
 
We completely restructure and greatly improve the efficiency of the Mathematica code developed in \cite{Chang:2013fba}.  Our algorithm can be outlined as follows.
\begin{enumerate}
    \item  List all the single-trace BPS words made out of the BPS letters up to a certain value of $n = n_*$.  Express each single-trace BPS word in the component form, and take into account the $\SU$ traceless condition by substituting away the $(N,N)$-entry.
    \item  List all the multi-trace BPS words in the component form up to $n = n_*$ by combining the single-trace BPS words.  Eliminate the linearly dependent multi-trace BPS words due to trace relations.  Let $V_Y$ denote the resulting vector space in a fixed-charge sector of $(J_L^3, J_R^3, q_j, Y)$, with $Y$ (denoting the homogeneous degree in $\Psi$) explicit for later convenience. 
    See Section~\ref{sec:review_coho} for the notations.
    \item  List all the single-graviton operators, express them in the component form, and then list all the multi-graviton operators by combining single-gravitons.  Let $W_Y$ denote the resulting vector space.
    \item  In the above steps, the presence of fermionic BPS letters means that the ordering of letters cannot be completely forgotten.  We fix an ordering on the BPS letters, and for each word commute the letters into the fixed ordering, while carefully keeping track of minus signs from fermionic statistics.
    \item  Act the supercharge $Q$ on $V_Y$, and compute bases for the spaces 
    \[ V_Y, ~~ QV_Y, ~~ {\rm span}(W_Y \cup QV_{Y-1}) \]
    by performing numerical QR decomposition in Julia.\footnote{We have been extremely careful with numerical stability.  In particular, the non-graviton cohomology classes corresponding to candidate black hole microstates discussed in Section~\ref{sec:microstates} have been verified with exact row reduction in Mathematica.
    }
    This gives matrix representations of the cochain complex and the embedding of ${\rm span}(W_Y \cup QV_{Y-1})$ into $V_Y$,
    \ie
    \begin{tikzcd}
    \cdots\arrow[r, "Q"]  & V_{Y-1}\arrow[r, "Q"]\arrow[rd,swap,"Q" ] & V_Y \arrow[r, "Q"] & V_{Y+1} \arrow[r, "Q"] & \cdots 
    \\
    & & {\rm span}(W_Y \cup QV_{Y-1}) \arrow[u,hook]\arrow[ru,swap,"Q=0"] &
    \end{tikzcd}
    \fe
    The representatives of all the cohomology classes can then be \emph{explicitly} constructed.
    \item
    For the purpose of counting, the dimensionality of the $Q$-cohomology is given by
    \[ \dim(V_Y) - \dim(QV_Y) - \dim(QV_{Y-1}), \] 
    and that restricted to the $Q$-cohomology with representatives of multi-graviton form is given by 
    \[ \dim({\rm span}(W_Y \cup QV_{Y-1})) - \dim(QV_{Y-1}). \]
\end{enumerate}

\subsection{Counting data and a giant-graviton-like feature}

\begin{figure}
    \centering
    \includegraphics[width=.45\textwidth]{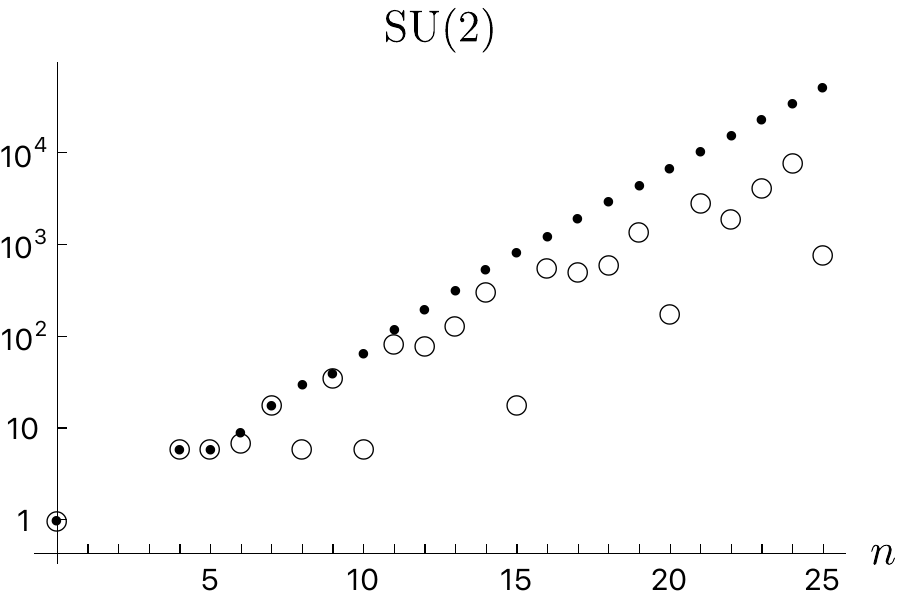} \quad \includegraphics[width=.45\textwidth]{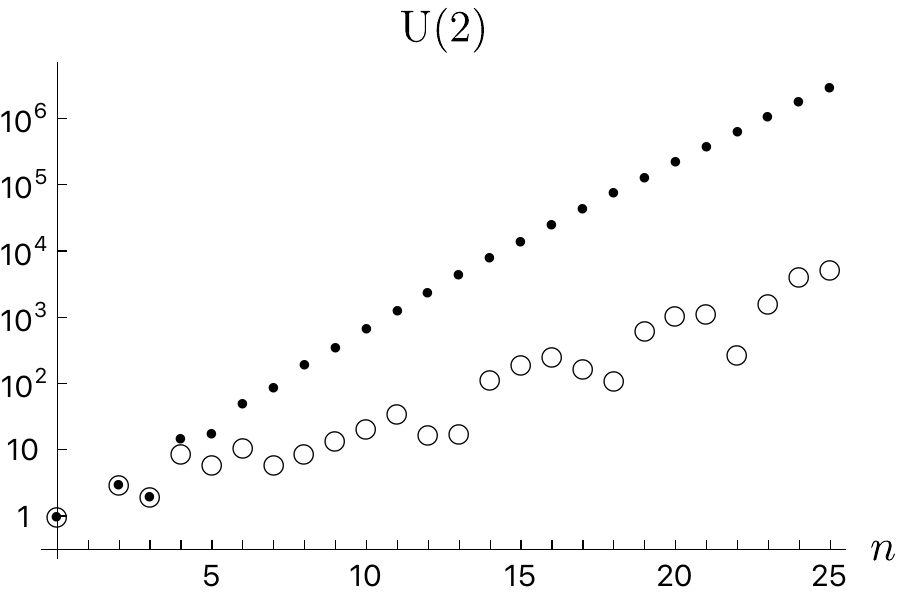}
    \\
    ~
    \\
    \includegraphics[width=.45\textwidth]{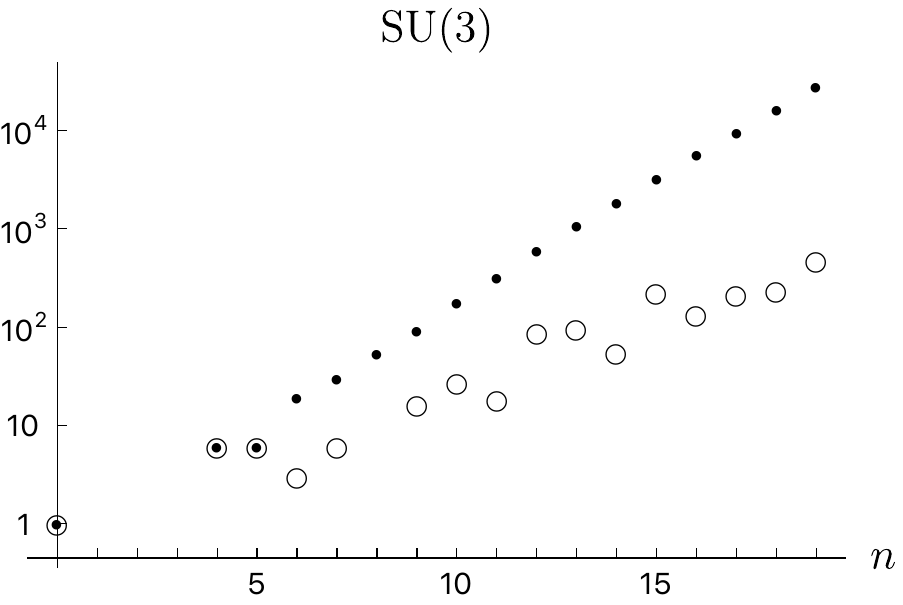} \quad \includegraphics[width=.45\textwidth]{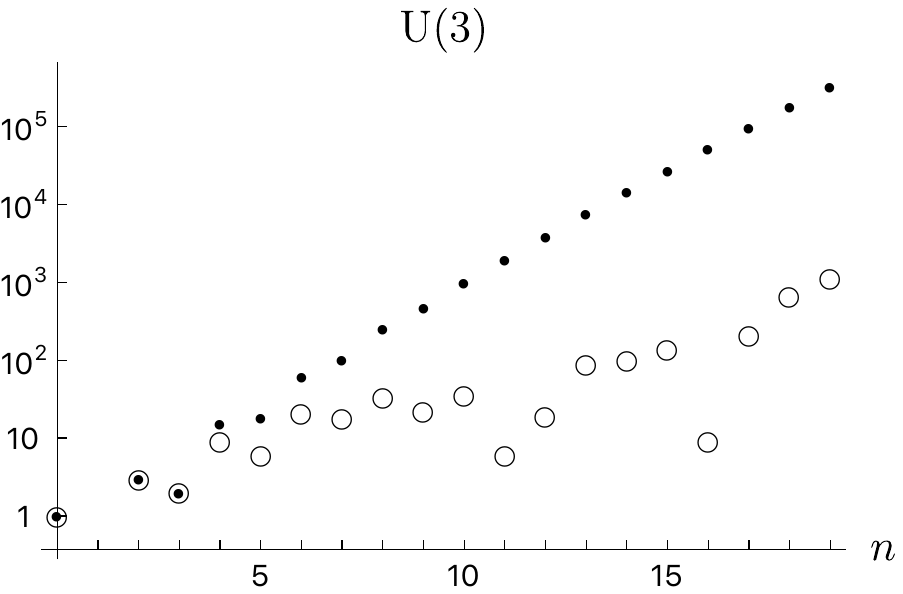}
    \\
    ~
    \\
    \includegraphics[width=.45\textwidth]{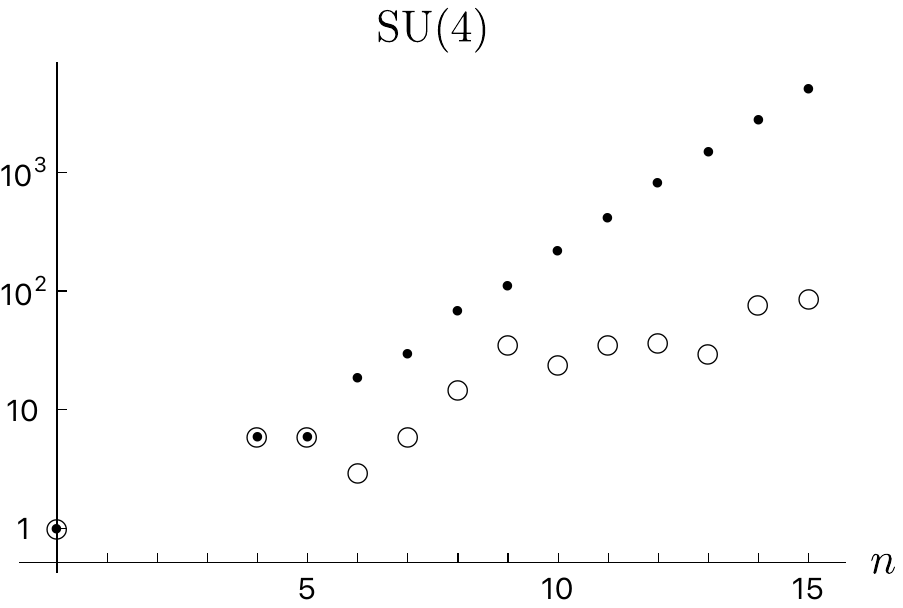} \quad \includegraphics[width=.45\textwidth]{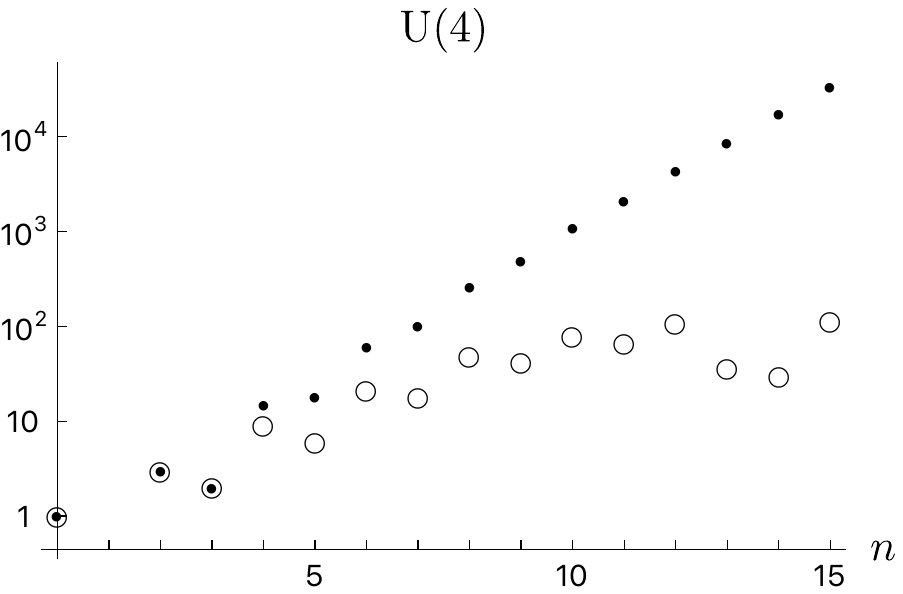}
    \caption{Logarithmic plot of the number of 1/16-BPS states ${\bf d}_N(n)$ (dots) and the absolute-valued index $d_N(n)$ (circles) at each $n$ for $\SU(N)$ and $\U(N)$ gauge groups, $N = 2, 3, 4$.}
    \label{fig:counting}
\end{figure}

We have performed the fully-refined cohomological enumeration comprehensively up to the values of $n$ indicated in Table~\ref{tab:n}. 
While the holographic duality concerns $\SU(N)$ gauge group, the index counting literature mostly considers $\U(N)$ gauge group.  We present the counting data of both for the convenience of the reader, even though the two are simply related by the contribution of a single decoupled D3-brane.
Figure~\ref{fig:counting} depicts the unrefined counting data.
The fully-refined counting data can be publicly accessed on \href{https://github.com/yinhslin/bps-counting}{https://github.com/yinhslin/bps-counting}, and will be continually updated as our computation progresses.

\begin{figure}
    \centering
    \includegraphics[width=.45\textwidth]{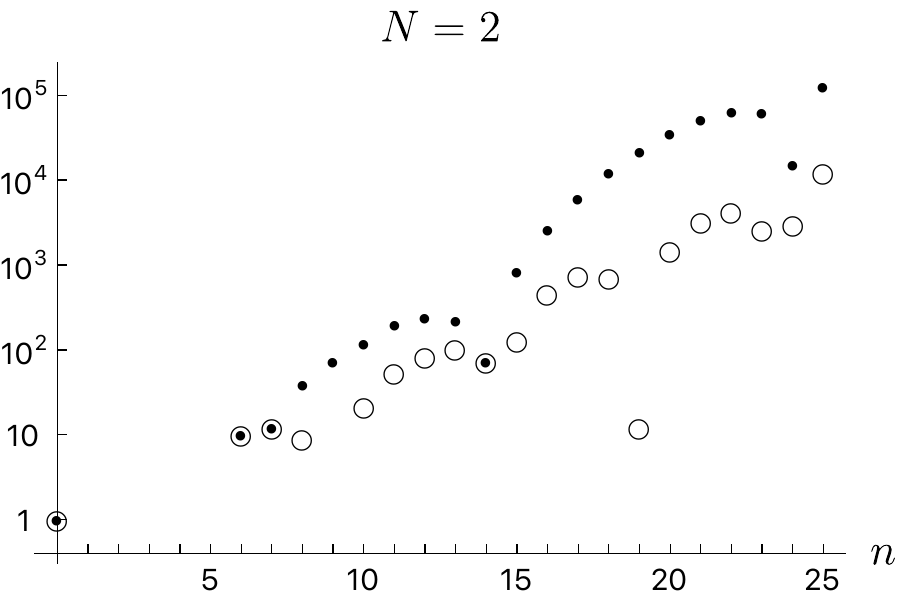}
    \\
    ~
    \\
    \includegraphics[width=.45\textwidth]{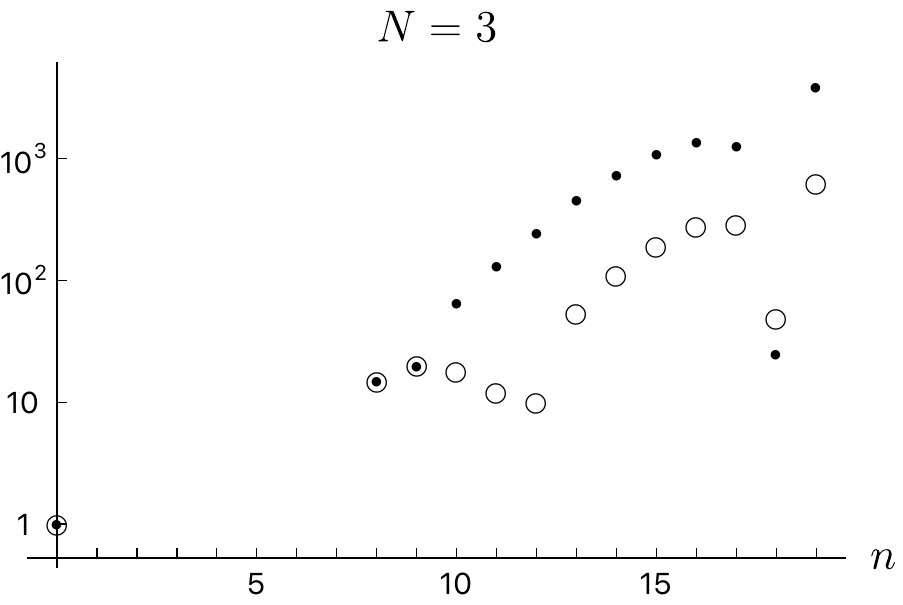}
    \\
    ~
    \\
    \includegraphics[width=.45\textwidth]{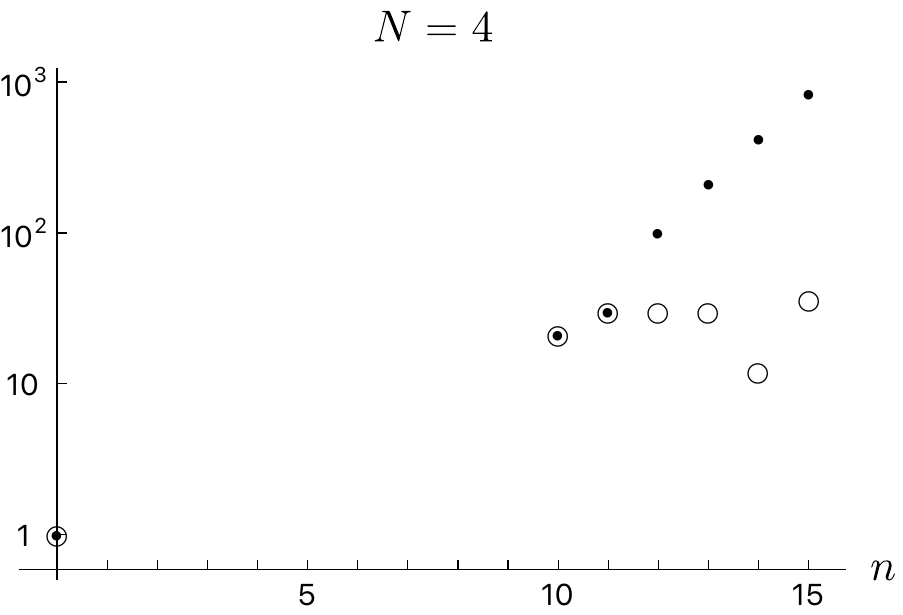}
    \caption{Logarithmic plot of the absolute-valued coefficients of $Z_N/Z_{N=\infty}$ (dots) and $I_N/I_{N=\infty}$ (circles), for $N = 2, 3, 4$.}
    \label{fig:giantgraviton}
\end{figure}

Inspired by the recent discovery of the giant graviton expansion \cite{Arai:2019xmp,Arai:2020qaj,Imamura:2021ytr,Gaiotto:2021xce,Lee:2022vig}, we divide the finite-$N$ BPS partition function (resp.\ index) by the infinite-$N$ one, and plot the coefficients at each $n$ in Figure~\ref{fig:giantgraviton}.  Note that the counting of 1/16-BPS states at finite-$N$ coincides with infinite-$N$ up to $n=2N+1$, as was explained by \cite{Murthy:2020rbd}.  A key signature of the giant graviton expansion is the presence of periodic ``dips'' indicating the contributions of giant gravitons.\footnote{We thank Nathan Benjamin and Ji Hoon Lee for a discussion.}  We see that not only the index but also the partition function exhibits such dips!  Such dips come from the sign-changes in the coefficients of $Z_{N=2}/Z_{N=\infty}$.  It would be fascinating if a giant graviton expansion of the partition function exists.  
All in all, we believe that the data we accumulated contains profound information about bulk quantum gravity.

The data underlying Figures~\ref{fig:counting} and~\ref{fig:giantgraviton} are tabulated in Tables~\ref{tab:2} through~\ref{tab:4}.

\begin{table}[t]
    \centering
    \begin{tabular}{|c|c|}
        \hline
        $N$ & Maximal $n$
        \\\hline\hline
        2 & 25
        \\
        3 & 19
        \\
        4 & 15
        \\
        \hline
    \end{tabular}
    \caption{The maximal $n$ of comprehensive cohomological enumeration for each $N$.}
    \label{tab:n}
\end{table}

\subsection{Black hole microstates}
\label{sec:microstates}

For $\SU(2)$ gauge group, we found that up to $n=25$, almost all cohomology elements have a representative that is of multi-graviton form, \emph{except} one element at $E=19/2$ and $n=24$ with the total number of derivatives given by $(\#\partial_{z^\pm}, \#\partial_{\theta_i}) = (0,0,4,4,4)$, as well as six elements at $E=10$ and $n=25$ with $(\#\partial_{z^\pm}, \#\partial_{\theta_i}) = (0,1,3,4,4)$ plus permutations.  
For $N=3,4$, every state is found to be of multi-graviton form, up to the values of $n$ indicated in Table~\ref{tab:n}.  The scarcity of candidates for black hole microstates at computationally-accessible charges explains the then-negative result of \cite{Chang:2013fba}.  Note that the enigma there was due to the lack of any \emph{candidates} for black hole microstates.  We are by no means suggesting that every state not of multi-graviton form must be a black hole microstate, but the discovery of viable candidates is reassuring.

For $\SU(2)$, the smallest value of $n=24$ at which a non-graviton state is found coincides with the location of the second dip of $Z_{N=2}/Z_{N=\infty}$ as shown in Figure~\ref{fig:giantgraviton}.  Could these non-graviton states have interpretations in terms of giant gravitons?\footnote{One important caveat is that the known giant gravitons have vanishing Lorentz spin, but the non-graviton states we found have nonzero Lorentz spin.  We thank Ji Hoon Lee for pointing this out.}

We stress that our search is comprehensive $n$-by-$n$, and not energy-by-energy.
By the BPS condition, states with a fixed $n$ have energies bounded below by $E \ge \lceil 2n/3 \rceil/2$ for $n>9$.  Our exhaustive search for $\SU(2)$ up to $n=25$ proves that there is no black hole microstate for $E < 9$.
By comparison, the expectation from the bulk side, valid at large $N$, is that black hole microstates should show up at energies $E \sim N^2$ or higher \cite{Kinney:2005ej}.  Hence, the actual lowest energy for $N=2$ is significantly higher than this bulk expectation.

Furthermore, while we have obtained explicit expressions for the space of representatives of each cohomology class, the precise 1/16-BPS operator must be annihilated by $Q^\dag$, and is hence a perturbative series in the coupling.
Demanding the annihilation by the one-loop $Q^\dag$ \eqref{Qdagger} will give the weak coupling of the 1/16-BPS operator, but it is unclear whether distinct properties of black hole microstates persist in the transition from strong to weak coupling.

\subsection{Outlook}

What are the salient features of black hole microstates when compared to multi-graviton states?  Are there linguistic rules governing the words and letters that underlie information-theoretic properties of black holes, such as chaos?  Compared to holographic descriptions of black holes in the canonical ensemble, such as the thermofield double \cite{Maldacena:2001kr}, our microcanonical data contains different information and may potentially allow the direct study of things like the eigenstate thermalization hypothesis \cite{deutsch1991quantum,Srednicki:1994mfb,rigol2008thermalization}.  We envision the computation performed in this paper to be the beginning of a ``black hole genome project'' that aims to provide key clues for these problems.  To this end, a more efficient way of enumerating 1/16-BPS states and black hole microstates is highly desired. The superconformal index admits an integral formula over (special) unitary matrices from the supersymmetric localization of the ${\cal N}=4$ SYM path integral. Could the BPS partition function also be computed by localization techniques?

While there is a closed form expression for the multi-graviton states as products of the single-graviton operators \eqref{eqn:graviton_op}, we do not have a simple way to write down the operators that represent the non-graviton cohomology classes. A special class of 1/16-BPS operators was written down in determinant form in 
\cite{Berkooz:2006wc,Berkooz:2008gc}. It would be interesting to investigate whether those determinant operators are of multi-graviton form, or whether they provide words for non-graviton cohomology classes.

Finally, the one-loop Beisert Hamiltonian in various subsectors has been explored in \cite{Harmark:2014mpa,Baiguera:2020jgy,Baiguera:2020mgk,Baiguera:2021hky}, and dubbed as the ``spin matrix theory" that describes the dynamics of near-BPS states. We have shown that the Beisert Hamiltonian takes a very simple form \eqref{Qdagger} and \eqref{eqn:H=QQ} in terms of the BPS superfield $\Psi$, which efficiently organizes all the BPS letters. This could help the study of the spin matrix theory in the largest subsector, the ${\rm psu}(1,2|3)$ subsector. On the bulk side, it has been proposed that the near-BPS states describe the near-horizon excitations of black holes, which are captured by the ${\cal N}=2$ Schwarzian theory \cite{Boruch:2022tno}. It would be fascinating to identify a Schwarzian sector of the spin matrix theory or the quantum mechanics of the all-loop Beisert Hamiltonian.

\section*{Acknowledgements}

We owe our gratitude to Nathan Benjamin and Ji Hoon Lee for a discussion that partially motivated this work, and to Xi Yin for insightful discussions and comments on the draft.
CC is partly supported by National Key R\&D Program of China (NO. 2020YFA0713000).
YL is supported by the Simons Collaboration Grant on the Non-Perturbative Bootstrap.  YL thanks New York University for its hospitality during the progression of this work.  The computations in this paper were run on the FASRC Cannon cluster supported by the FAS Division of Science Research Computing Group at Harvard University.

\begin{table}
    \centering
    \[
    \begin{array}{|c||c|c||c|c||c|c|}
        \hline
        n & {\bf d}_{\SU(2)} & d_{\SU(2)} & {\bf d}_{\U(2)} & d_{\U(2)} & Z_{N=2}/Z_{N=\infty}|_{t^n} & I_{N=2}/I_{N=\infty}|_{t^n} \\\hline\hline
        0 & 1 & 1 & 1 & 1 & 1 & 1 \\\hline
        1 & 0 & 0 & 0 & 0 & 0 & 0 \\\hline
        2 & 0 & 0 & 3 & 3 & 0 & 0 \\\hline
        3 & 0 & 0 & 2 & -2 & 0 & 0 \\\hline
        4 & 6 & 6 & 15 & 9 & 0 & 0 \\\hline
        5 & 6 & -6 & 18 & -6 & 0 & 0 \\\hline
        6 & 9 & -7 & 51 & 11 & -10 & -10 \\\hline
        7 & 18 & 18 & 90 & -6 & -12 & 12 \\\hline
        8 & 30 & 6 & 195 & 9 & -39 & -9 \\\hline
        9 & 40 & -36 & 362 & 14 & -72 & 0 \\\hline
        10 & 66 & 6 & 699 & -21 & -117 & 21 \\\hline
        11 & 120 & 84 & 1308 & 36 & -198 & -54 \\\hline
        12 & 198 & -80 & 2431 & -17 & -237 & 83 \\\hline
        13 & 324 & -132 & 4434 & -18 & -222 & -102 \\\hline
        14 & 537 & 309 & 8046 & 114 & 72 & 72 \\\hline
        15 & 822 & -18 & 14346 & -194 & 840 & 128 \\\hline
        16 & 1257 & -567 & 25434 & 258 & 2577 & -459 \\\hline
        17 & 1944 & 516 & 44544 & -168 & 6084 & 744 \\\hline
        18 & 2959 & 613 & 77442 & -112 & 12067 & -697 \\\hline
        19 & 4476 & -1392 & 133386 & 630 & 21660 & 12 \\\hline
        20 & 6834 & -180 & 228021 & -1089 & 35166 & 1440 \\\hline
        21 & 10352 & 2884 & 386898 & 1130 & 51136 & -3240 \\\hline
        22 & 15540 & -1926 & 651843 & -273 & 64368 & 4182 \\\hline
        23 & 23406 & -4242 & 1091004 & -1632 & 61440 & -2580 \\\hline
        24 & 35076 & 7890 & 1814578 & 4104 & 15129 & -2971 \\\hline
        25 & 52020 & 792 & 2999724 & -5364 & -124884 & 12132 \\\hline
    \end{array}
    \]
    \caption{Counting of 1/16-BPS states for $\SU(2)$ and $\U(2)$ gauge groups, and the coefficients of $Z_{N=2}/Z_{N=\infty}$ and $I_{N=2}/I_{N=\infty}$.}
    \label{tab:2}
\end{table}

\begin{table}
    \centering
    \[
    \begin{array}{|c||c|c||c|c||c|c|}
        \hline
        n & {\bf d}_{\SU(3)} & d_{\SU(3)} & {\bf d}_{\U(3)} & d_{\U(3)} & Z_{N=3}/Z_{N=\infty}|_{t^n} & I_{N=3}/I_{N=\infty}|_{t^n} \\\hline\hline
        0 & 1 & 1 & 1 & 1 & 1 & 1 \\\hline
        1 & 0 & 0 & 0 & 0 & 0 & 0 \\\hline
        2 & 0 & 0 & 3 & 3 & 0 & 0 \\\hline
        3 & 0 & 0 & 2 & -2 & 0 & 0 \\\hline
        4 & 6 & 6 & 15 & 9 & 0 & 0 \\\hline
        5 & 6 & -6 & 18 & -6 & 0 & 0 \\\hline
        6 & 19 & 3 & 61 & 21 & 0 & 0 \\\hline
        7 & 30 & 6 & 102 & -18 & 0 & 0 \\\hline
        8 & 54 & 0 & 249 & 33 & -15 & -15 \\\hline
        9 & 92 & -16 & 470 & -22 & -20 & 20 \\\hline
        10 & 177 & 27 & 996 & 36 & -66 & -18 \\\hline
        11 & 318 & 18 & 1938 & 6 & -132 & 12 \\\hline
        12 & 595 & -87 & 3865 & -19 & -246 & 10 \\\hline
        13 & 1068 & 96 & 7458 & 90 & -462 & -54 \\\hline
        14 & 1854 & 54 & 14373 & -99 & -735 & 111 \\\hline
        15 & 3234 & -222 & 27258 & 138 & -1106 & -190 \\\hline
        16 & 5610 & 132 & 51339 & -9 & -1371 & 279 \\\hline
        17 & 9558 & 210 & 95562 & -210 & -1284 & -288 \\\hline
        18 & 16329 & -235 & 176594 & 672 & 25 & 49 \\\hline
        19 & 27612 & -468 & 323208 & -1116 & 3870 & 630 \\\hline
    \end{array}
    \]
    \caption{Counting of 1/16-BPS states for $\SU(3)$ and $\U(3)$ gauge groups, and the coefficients of $Z_{N=3}/Z_{N=\infty}$ and $I_{N=3}/I_{N=\infty}$.}
    \label{tab:3}
\end{table}

\begin{table}
    \centering
    \[
    \begin{array}{|c||c|c||c|c||c|c|}
        \hline
        n & {\bf d}_{\SU(4)} & d_{\SU(4)} & {\bf d}_{\U(4)} & d_{\U(4)} & Z_{N=4}/Z_{N=\infty}|_{t^n} & I_{N=4}/I_{N=\infty}|_{t^n} \\\hline\hline
        0 & 1 & 1 & 1 & 1 & 1 & 1 \\\hline
        1 & 0 & 0 & 0 & 0 & 0 & 0 \\\hline
        2 & 0 & 0 & 3 & 3 & 0 & 0 \\\hline
        3 & 0 & 0 & 2 & -2 & 0 & 0 \\\hline
        4 & 6 & 6 & 15 & 9 & 0 & 0 \\\hline
        5 & 6 & -6 & 18 & -6 & 0 & 0 \\\hline
        6 & 19 & 3 & 61 & 21 & 0 & 0 \\\hline
        7 & 30 & 6 & 102 & -18 & 0 & 0 \\\hline
        8 & 69 & 15 & 264 & 48 & 0 & 0 \\\hline
        9 & 112 & -36 & 490 & -42 & 0 & 0 \\\hline
        10 & 222 & 24 & 1086 & 78 & -21 & -21 \\\hline
        11 & 420 & 36 & 2130 & -66 & -30 & 30 \\\hline
        12 & 831 & -37 & 4411 & 107 & -100 & -30 \\\hline
        13 & 1530 & -30 & 8676 & -36 & -210 & 30 \\\hline
        14 & 2844 & 78 & 17280 & 30 & -420 & -12 \\\hline
        15 & 5220 & 88 & 33670 & 114 & -832 & -36 \\\hline
    \end{array}
    \]
    \caption{Counting of 1/16-BPS states for $\SU(4)$ and $\U(4)$ gauge groups, and the coefficients of $Z_{N=4}/Z_{N=\infty}$ and $I_{N=4}/I_{N=\infty}$.}
    \label{tab:4}
\end{table}

\newpage

\bibliography{refs}
\bibliographystyle{JHEP}

\end{document}